\documentclass[aps,pra,twocolumn,superscriptaddress,raggedbottom,showpacs,floatfix,longbibliography]{revtex4-1}
\usepackage{amsmath,amsfonts,amssymb,amsthm,graphics}
\usepackage[next]{inputenc}
\usepackage[colorlinks=true,citecolor=blue,linkcolor=blue]{hyperref}
\usepackage{bbm}
\usepackage{bbold}
\usepackage{booktabs}
\usepackage{multirow}
\usepackage{hhline}
\usepackage{amssymb}
\usepackage{comment}
\usepackage{float}
\usepackage{latexsym}
\usepackage{bm}
\usepackage[dvipsnames]{xcolor}
\usepackage{graphicx}
\usepackage{epstopdf}

\begin{document}
\renewcommand{\vec}{\mathbf}
\renewcommand{\Re}{\mathop{\mathrm{Re}}\nolimits}
\renewcommand{\Im}{\mathop{\mathrm{Im}}\nolimits}
\renewcommand{\phi}{\varphi}

\title{Kinetic theory of dark solitons with tunable friction}
\author{Hilary M. Hurst}
\affiliation{Joint Quantum Institute and Condensed Matter Theory Center, Department of Physics, University of Maryland, College Park, Maryland 20742-4111, USA}
\author{Dmitry K. Efimkin}
\affiliation{The Center for Complex Quantum Systems, The University of Texas at Austin, Austin, Texas 78712-1192, USA}
\author{I. B. Spielman}
\affiliation{Joint Quantum Institute, National Institute of Standards and Technology, and University of Maryland, Gaithersburg, Maryland, 20899, USA}
\author{Victor Galitski}
\affiliation{Joint Quantum Institute and Condensed Matter Theory Center, Department of Physics, University of Maryland, College Park, Maryland 20742-4111, USA}

\begin{abstract}
We study controllable friction in a system consisting of a dark soliton in a one-dimensional Bose-Einstein condensate coupled to a non-interacting Fermi gas. The fermions act as impurity atoms, not part of the original condensate, that scatter off of the soliton. We study semi-classical dynamics of the dark soliton, a particle-like object with negative mass, and calculate its friction coefficient. Surprisingly, it depends periodically on the ratio of interspecies (impurity-condensate) to intraspecies (condensate-condensate) interaction strengths. By tuning this ratio, one can access a regime where the friction coefficient vanishes. We develop a general theory of stochastic dynamics for negative mass objects and find that their dynamics are drastically different from their positive mass counterparts - they do not undergo Brownian motion. From the exact phase space probability distribution function (i.e. in position and velocity), we find that both the trajectory and lifetime of the soliton are altered by friction, and the soliton can only undergo Brownian motion in the presence of friction and a confining potential. These results agree qualitatively with experimental observations by Aycock, et. al. (PNAS, 2017) in a similar system with bosonic impurity scatterers. 
\end{abstract}

\maketitle

\section{Introduction\label{Sec:Intro}} Solitons are common in physical systems, from water waves to optical pulses. These particle-like excitations propagate without changing their shape; their remarkable stability is related to the integrability of the underlying nonlinear equations describing them. Advances in ultracold quantum gases have lead to experiments with precise control over the manipulation and creation of solitons. Dark solitons, associated with a dip in density, and bright solitons, associated with a bump in density, have been experimentally observed~\cite{Burger1999,Denschlag2000,Khaykovich2002,Strecker2002,Yefsah2013}. In the near future, more complicated structures such as vector solitons and magnetic solitons could be realized in ultracold gases. Spinor condensates, Bose-Fermi mixtures, and binary condensates have all been proposed as systems for exploring exotic soliton physics~\cite{Becker2008,Ieda2004,Karpiuk2004,Tercas2013,Tercas2014,Qu2016}. Furthermore, solitons in quantum gases are close relatives of magnetic solitons in solid state systems, such as domain walls and skyrmions, which have been proposed as candidates for information storage~\cite{Braun1996,Wong2010,Kim2014,Kim2015,Psaroudaki2016}. Theoretical understanding of how dissipation and noise affects solitons is essential for their incorporation in future technologies.\\
\indent Matter-wave dark solitons in Bose-Einstein condensates (BEC) are ideal probes of both classical and quantum dynamics. Their heavy mass and localized nature allow their dynamics to be understood classically. However, they are also highly sensitive to dimensionality and background fluctuations~\cite{Anglin2008}. Recent years have seen renewed theoretical interest in the effects of friction and dissipation on solitons, which can greatly affect their lifetime and stability~\cite{Fedichev1999,Efimkin2015,Mcdonald2016,Efimkin2016,Jackson2007,Cockburn2011,Muryshev2002,Sinha2006}. Additionally, the diffusion coefficient of a soliton was recently measured experimentally for the first time~\cite{Aycock2016}.\\
\indent In the theoretical literature, dissipation in solitonic systems has been studied by including a trap or introducing additional dimensions~\cite{Fedichev1999,Ivanov1990,Parker2003,Parker2010}. In higher dimensions, solitons are unstable and experimental systems must be close to one-dimensional (1D) in order to observe them. The friction coefficient of the soliton is related to the reflection coefficient of excitations scattering off of it~\cite{Fedichev1999,Busch2001,Jackson2007,Muryshev2002}. In isolated Bose gases at non-zero temperature, these scatterers would be the Bogoliubov quasiparticles of the original condensate, and their reflection coefficient can be calculated for various geometries. However, in a 1D scattering theory and without a trap the Bogoliubov excitations are reflectionless, and therefore do not cause Ohmic friction~\cite{Muryshev2002,Efimkin2016}.  It was recently shown that non-Ohmic friction can still occur in these systems by accounting for non-Markovian effects~\cite{Efimkin2016}. However, experimental system are necessarily quasi-1D, therefore both Ohmic friction and non-Markovian friction are present.\\
\indent In this work we consider an alternative way to induce Ohmic friction in a 1D system consisting of a condensate, with a dark soliton, coupled to a non-interacting cloud of fermionic ``impurity" atoms. This is similar to the setup employed in Ref.~\cite{Aycock2016}. We present three main results: First, the system \emph{with} impurities can be tuned to have zero Ohmic friction, based only on the ratio of interspecies (impurity-condensate) to intraspecies (condensate-condensate) interaction strengths. Secondly, in contrast to objects with positive mass, for a negative mass object such as a dark soliton there is no diffusion in a meaningful sense in free space. We show that the soliton undergoes only ballistic motion due to the fact that friction increases its speed, providing an anti-damping force. Third, in the presence of Ohmic friction \emph{and} an external potential, the dark soliton can undergo diffusion or Brownian motion, characterized by a mean squared displacement that grows linearly in time, $\langle \bar{x}^2\rangle \propto Dt$. In this case, the diffusion coefficient is $D\propto \gamma/\omega^2$, where $\gamma$ is the friction coefficient and $\omega$ is the frequency of harmonic confinement. The diffusion coefficient is \emph{proportional} to the amount of friction in the system, in contrast to the usual case where $D \propto 1/\gamma$~\cite{Einstein1905}. Dark solitons provide an ideal experimental testbed for the mechanism of trap-induced Brownian motion.\\
\indent The paper is structured as follows: In Sec.~\ref{Sec:Model} we outline the model of a dark soliton in quasi-1D BEC in the presence of non-interacting fermions. In Sec.~\ref{Sec:Impurity} we discuss the single-particle scattering properties of the fermions in the presence of the soliton, which acts as a potential well for the fermions. Sec.~\ref{Sec:Kinetic} is devoted to kinetic theory, where we derive two essential equations: the microscopic expression for the friction coefficient, and the kinetic equation for the soliton probability distribution function (PDF), which can be calculated exactly. In Sec.~\ref{Sec:Dark}, we use the PDF to calculate the soliton's average position and variance in position. We show that Brownian motion only occurs in the presence of an external trap and calculate the diffusion coefficient. We use the PDF again in Sec.~\ref{Sec:Soliton} to define and calculate the soliton lifetime. Finally, in Sec.~\ref{Sec:Discussion} we discuss possible experimental implementations of our proposal and conclude. Technical details of the calculations are left to the Appendices.
\section{Model\label{Sec:Model}} We consider a quasi-1D bosonic superfluid interacting with a Fermi gas in an external potential. The proposed creation and manipulation of solitons requires a highly elongated geometry with confinement frequency $\omega_{\mathrm{c/i},x} \ll \omega_{\mathrm{c/i}\perp}$, where the subscript c denotes the bosons that make up the condensate, subscript i denotes fermionic impurity atoms, and $\omega_{\mathrm{c/i},x}$ and $\omega_{\mathrm{c/i}\perp}$ denote the confinement frequencies for the elongated and transverse directions, respectively.\\
\indent A 1D theory is sufficient to describe the quasi-1D system provided that the transverse confinement is tight enough that transverse degrees of freedom can be eliminated, conditions which we enumerate below. Under these conditions, the system is described by the 1D Hamiltonian $\hat{H} = \hat{H}_{\rm c} + \hat{H}_{\rm i} + \hat{H}_{\rm int}$,
\begin{align}
&\hat{H}_{\rm c} = \int dx~\frac{\hbar^2}{2m_{\rm c}}\nabla\hat{\phi}^\dagger\nabla\hat{\phi} + U(x)\hat{\phi}^\dagger\hat{\phi} + \frac{g}{2}\hat{\phi}^\dagger\hat{\phi}^\dagger\hat{\phi}\hat{\phi}\label{eqn:H1}\\
&\hat{H}_{\rm i} = \int dx~\frac{\hbar^2}{2m_{\rm i}}\nabla\hat{\psi}^\dagger\nabla\hat{\psi} + U(x)\hat{\psi}^\dagger\hat{\psi}\label{eqn:H2}\\
&\hat{H}_{\rm int} = \int dx~g'\hat{\psi}^\dagger\hat{\phi}^\dagger\hat{\phi}\hat{\psi}\label{eqn:H3},
\end{align}
where $U(x)$ is an external potential, $\hbar$ is Planck's constant and $m_\mathrm{c}$ and $m_\mathrm{i}$ denote the masses of the condensate and impurity atoms, respectively. The field operators are denoted $\hat{\phi}$ for bosons and $\hat{\psi}$ for fermions. By integrating over transverse degrees of freedom, the 1D interaction strengths are given by the well known expressions $g = 2\hbar\omega_{\mathrm{c}\perp}a_\mathrm{cc}$ and $g' = 2\hbar\sqrt{\omega_{\mathrm{c}\perp}\omega_{i\perp}}a_\mathrm{ci}$ where $a_\mathrm{cc}$ and $a_\mathrm{ci}$ denote the three-dimensional intraspecies (boson-boson) and interspecies (boson-fermion) scattering lengths~\cite{Olshanii1998,Imambekov2006}.\\
\indent At very low temperatures the bosons undergo Bose-Einstein condensation. Provided that the bosons are weakly interacting ($gn_{\rm c}\ll 1$ where $n_{\rm c}$ is the density), we can make the mean-field approximation $\langle\hat{\phi}\rangle \rightarrow \phi_0$. The field $\phi_0$ denotes the macroscopic wavefunction of the condensate, which obeys the Gross-Pitaevskii equation (GPE) 
\begin{equation}
    i\hbar\frac{\partial\phi_0}{\partial t} = -\frac{\hbar^2}{2m_\mathrm{c}}\frac{\partial^2\phi_0}{\partial x^2} + U(x)\phi_0 + g|\phi_0|^2\phi_0 + g'n_{\rm i}\phi_0\label{eqn:condensate},
\end{equation}
where $n_{\rm i} = \langle \hat{\psi}^\dagger\hat{\psi}\rangle$ is the impurity density.\\
\indent The condensate profile then appears as an external potential $V(x) = |\phi_0(x)|^2$ for the impurity atoms, which we treat using a single-particle model. The single particle wavefunction of the fermions, denoted $\psi$, obeys the Schr\"{o}dinger equation
\begin{equation}
    i\hbar\frac{\partial\psi}{\partial t} = -\frac{\hbar^2}{2m_\mathrm{i}}\frac{\partial^2\psi}{\partial x^2} + U(x)\psi +  g'|\phi_0|^2\psi. \label{eqn:imp}
\end{equation}
\indent Hamiltonian equations~\eqref{eqn:H1}-\eqref{eqn:H3} apply to both bosonic and fermionic impurities; in this work we consider the latter. Such Bose-Fermi mixtures have been realized experimentally and have been shown to be stable in quasi-1D~\cite{Lai1971,Imambekov2006,Truscott2001,Schreck2001,Hadzibabic2002,Goldwin2004,Roy2017}. In order for our 1D theory to be applicable, the system must be in the quasi-1D regime. This corresponds to the condition $\mu_\mathrm{c} \ll \hbar \omega_{\mathrm{c}\perp}$ for the condensate and $\mu_\mathrm{i}\ll \hbar\omega_{\mathrm{i\perp}}$ for the impurity atoms, where $\mu_\mathrm{c,i}$ is the chemical potential of the condensate and impurities, respectively.\\
\indent In the microscopic theory, the harmonic potential $U(x)$ is assumed to be sufficiently shallow such that $l_{\rm t} \gg \xi$, where $l_{\rm t} = \sqrt{\hbar/m_{\rm c}\omega_{\mathrm{c},x}}$ is the effective length scale of the trap and $\xi$ is the healing length of the condensate. Therefore, the trap only weakly affects the solutions to equations~\eqref{eqn:condensate} and~\eqref{eqn:imp} and we set $U(x) = 0$ in the following.  The background (Thomas-Fermi) confining potential provided by the BEC in a trap will however be important when we consider the semiclassical dynamics of the soliton in later sections.\\
\indent Under the assumption $U(x) = 0$, equation~\eqref{eqn:condensate} is known to have dark soliton solutions of the form $\phi_0(x,t) = \tilde{\phi}_0(x-v_\mathrm{s}t)e^{-i\mu_\mathrm{c}t/\hbar}$, with 
\begin{equation}
    \tilde{\phi}_0(x-v_\mathrm{s}t) = \sqrt{n_\mathrm{c}}\left[i\frac{v_\mathrm{s}}{c} +\gamma_\mathrm{s}\tanh\left(\gamma_\mathrm{s}\frac{x-v_\mathrm{s}t}{\sqrt{2}\xi}\right)\right],
    \label{eqn:soliton}
\end{equation}
where $v_\mathrm{s}$ denotes the soliton velocity, $n_\mathrm{c}$ is the density of the condensate as $x \rightarrow \pm \infty$, $c = \sqrt{\mu_\mathrm{c}/m_\mathrm{c}}$ is the speed of sound in the condensate with chemical potential $\mu_\mathrm{c} = gn_\mathrm{c}$, $\gamma_\mathrm{s}^2 = 1-v_\mathrm{s}^2/c^2$, and $\xi = \hbar/\sqrt{2} m_\mathrm{c}c$ is the condensate healing length. We have also neglected the last term in equation~\eqref{eqn:condensate}, $\propto g'n_{\rm i}$, because at very low densities the impurity atoms do not impede soliton creation and it is safe to assume the typical dark soliton profile for the condensate wavefunction~\cite{Aycock2016}.\\
\indent We see from equation~\ref{eqn:imp} and~\eqref{eqn:soliton} that the dark soliton creates a potential well for the impurity atoms. First, we analyze the single particle scattering properties of the impurities due to the soliton well. In our further analysis we treat the soliton as a classical particle interacting with a bath of fermionic quantum scatterers, similar to the problem of a heavy particle moving through a gas of much lighter particles~\cite{Lifschitz1983}.
\begin{figure}[t!]
\centering
\includegraphics[width=\columnwidth]{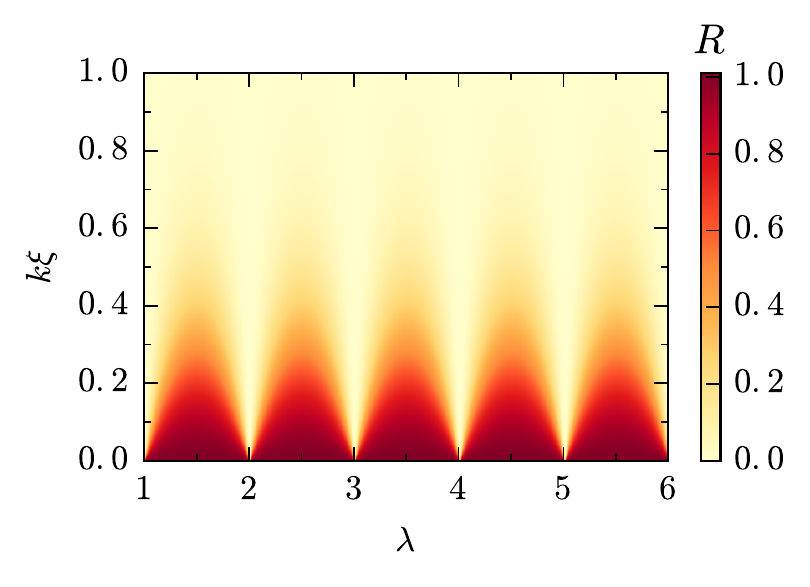}
\caption{(color online)~The reflection coefficient $R(k,\lambda)$ as a function of $\lambda$ where $\lambda(\lambda-1) = 2m_\mathrm{i}g'/m_\mathrm{c}g$. $R(k,\lambda)$ is strongly peaked at $k\approx 0$ and is periodic as a function of $\lambda$. When $\lambda$ is an integer $R(k,\lambda)$ is exactly zero.\label{Fig:RvsL}}
\end{figure}
\section{Impurity scattering\label{Sec:Impurity}} We can rewrite equation~\eqref{eqn:imp} in the frame co-moving with the soliton by making the variable transformation $z = \gamma_\mathrm{s}(x-v_\mathrm{s}t)/\sqrt{2}\xi$, $\psi(x,t) = e^{-iEt/\hbar}e^{ik_{\rm s}z}\psi(z)$ where $k_{\rm s} = m_{\rm i}v_{\rm s}/\hbar$. This gives the following time-independent Schr\"{o}dinger equation for impurity atoms, 
\begin{equation}
    \frac{\partial^2\psi(z)}{\partial z^2} + \left[\frac{\lambda(\lambda -1)}{\cosh^2z} + k^2\right]\psi(z) = 0, \label{eqn:PoschlTeller}
\end{equation}
where
\begin{equation}
    \lambda(\lambda-1) = \frac{2m_\mathrm{i}g'}{m_\mathrm{c}g} \mbox{~;~} k^2 = \frac{4m_\mathrm{i}\xi^2}{\hbar^2\gamma_\mathrm{s}^2}\left(E+\frac{m_\mathrm{i}v_\mathrm{s}^2}{2}-g'n_\mathrm{c}\right).
\end{equation}
The potential in equation~\eqref{eqn:PoschlTeller} is known as the P\"{o}schl-Teller potential, whose solutions are known in closed form %of hypergeometric functions
and has been widely studied in the context of supersymmetric quantum mechanics~\cite{Poschl1933,Frank1984,Guerrero2010,Alhassid1983,Ccevik2016,Lifschitz1983,Shaukat2017}. The reflection coefficient of scattering states is
\begin{equation}
    R(k,\lambda) = \frac{1-\cos(2\pi\lambda)}{\cosh(2\pi k)-\cos(2\pi\lambda)}.
    \label{eqn:Reflection}
\end{equation}
$R(k,\lambda)$ vs. $\lambda$ is shown in Figure \ref{Fig:RvsL}. Furthermore, $R(k,\lambda) = 0$ when $\lambda$ takes integer values, thus the soliton can become reflectionless to the impurities. The soliton potential well can also have bound states. The total number of bound states is the largest positive integer $j < \lambda-1$.
%also has a number of bound states, $j_\mathrm{bound} < \lambda -1$, where $j_\mathrm{bound} = 0,1,2,\ldots$ is an integer.
One or two fermionic impurities may occupy each bound state in the soliton core, and these bound particles affect the phase shift of scattered impurities. The effect of bound states and scattering state phase shifts are taken into account when we calculate the chemical potential of the fermions in Appendix A.
\section{Kinetic Theory of Dark Solitons\label{Sec:Kinetic}} The energy of the soliton texture in equation~\eqref{eqn:soliton} is calculated by subtracting the uniform background, and it is given by  
\begin{equation}
    E(v_\mathrm{s}) = \frac{4\hbar n_\mathrm{c}c}{3}\left(1-\frac{v_\mathrm{s}^2}{c^2}\right)^{3/2}\approx \frac{4\hbar n_\mathrm{c}c}{3}-\frac{M v_\mathrm{s}^2}{2}.
    \label{eqn:energy}
\end{equation}
Here we expanded $E(v_{\rm s})$ under the condition $v_{\rm s} \ll c$. The soliton is effectively a particle with negative mass of magnitude $M = 4\hbar n_\mathrm{c}/c = 4\sqrt{2}n_\mathrm{c}\xi m_\mathrm{c}$. The soliton is heavy compared to a single atom, $m_{\rm c} \ll M$, with width $\xi/\gamma_{\rm s}$. The heavy mass and localized nature of the dark soliton justifies the following classical treatment of its dynamics~\cite{Konotop2004}.\\
\indent We note here that $M$ is often called the ``inertial mass", whereas one can also define the ``gravitational mass" ($M_{\rm g}$) of a soliton, which is also negative. Gravitational mass is the missing mass of the atoms in the soliton core, given by integrating over the soliton density with the uniform background subtracted, $-M_{\rm g} = m_{\rm c}\int dx (n_{\rm s} - n_{\rm c})$, where $n_{\rm s}(x,t) = |\tilde{\phi}_0(x-v_st)|^2$ from equation~\eqref{eqn:soliton}. For a stationary soliton, $M_{\rm g} = 2\sqrt{2}n_{\rm c}\xi m_\mathrm{c}$ and $M = 2M_{\rm g}$. In the case of harmonic confinement the soliton has gravitational potential energy $U(x_{\rm s}) = -M_{\rm g}\omega_{\mathrm{c},x}^2x_{\rm s}^2/2$. This distinction is important for the soliton's classical equation of motion, given by $-M\ddot{x} = M_{\rm g}\omega^2_{\mathrm{c},x}x$. Dark solitons are quite stable in a harmonic trap and oscillate as a classical particle would, with effective frequency $\omega = \omega_{\mathrm{c},x}/\sqrt{2}$~\cite{Busch2000,Konotop2004,Weller2008}.\\
\indent In order to describe the diffusive behavior of solitons, we need to understand how a soliton will deviate from the average trajectory computed for many solitons. Soliton dynamics can be most readily examined by understanding their probability distribution function (PDF) $f(t,x_{\rm s}(t),v_{\rm s}(t))$. The soliton PDF obeys the kinetic equation 
\begin{equation}
    \frac{\partial f}{\partial t} + v_\mathrm{s}\frac{\partial f}{\partial x_\mathrm{s}} + \dot{v}_\mathrm{s}\frac{\partial f}{\partial v_\mathrm{s}}= \mathcal{I}\left[f\right],
    \label{eqn:kineticeqn}
\end{equation}
where the collision integral $\mathcal{I}\left[f\right]$ accounts for scattering of fermionic impurities off of the soliton. As the soliton is much heavier than the fermions, $M \gg m_\mathrm{i}$, the transferred momentum to and from the soliton due to collisions is small. This assumption results in a collision integral of Fokker-Plank form 
\begin{equation}
\mathcal{I}\left[f\right] = \frac{\partial}{\partial v_\mathrm{s}}\left(-A_{v_s}f + \frac{\partial}{\partial v_\mathrm{s}}\left[B_{v_s} f\right]\right),
\label{eqn:collisionint}
\end{equation}
which we derive in Appendix B. The transport coefficients $A_{v_s}$ and $B_{v_s}$ account for drift and diffusion of the distribution and are given by 
\begin{align}
A_{v_{\rm s}} &= -\frac{2\hbar}{M}\sum_k k R_{k,\lambda}\left|\frac{\partial \epsilon_k}{\hbar \partial k}\right| n_\mathrm{F}(\epsilon_{k+k_\mathrm{s}})\left[1-n_\mathrm{F}(\epsilon_{-k+k_\mathrm{s}})\right].
\label{eqn:A} \\
B_{v_{\rm s}} &= \frac{2\hbar^2}{M^2}\sum_k k^2R_{k,\lambda}\left|\frac{\partial \epsilon_k}{\hbar \partial k}\right| n_\mathrm{F}(\epsilon_{k+k_\mathrm{s}})\left[1-n_\mathrm{F}(\epsilon_{-k+k_\mathrm{s}})\right].\label{eqn:B}
\end{align}
Where $n_\mathrm{F}(\epsilon_k)$ is the Fermi-Dirac distribution for the impurity atoms, which is shifted by $k_\mathrm{s}$ because we calculated the reflection coefficient $R(k,\lambda)$ in frame co-moving with the soliton. The impurities have the usual dispersion relation $\epsilon_k = \hbar^2k^2/2m_\mathrm{i}$, and the last term $\left[1-n_\mathrm{F}(\epsilon_{-k})\right]$ accounts for Pauli-blocking effects on impurity scattering. \\
\indent To the lowest order in $k_\mathrm{s}$, we expand $n_{\rm F}(\epsilon_{k+k_{\rm s}})$ to find $A_{v_\mathrm{s}} = \gamma v_\mathrm{s}/M$ and $B_{v_{\rm s}} = \gamma k_{\rm B}T/M^2$, where $\gamma$ is given by 
\begin{equation}
    \gamma = \frac{2\hbar^2}{k_\mathrm{B}T}\sum_{k}k^2R_{k,\lambda}\left|\frac{\partial \epsilon_k}{\hbar\partial k}\right|n_\mathrm{F}(\epsilon_k)\left[1-n_\mathrm{F}(\epsilon_{-k})\right].
    \label{eqn:gamma}
\end{equation} 
\indent This exact expression shows that $A_{v_s}$ and $B_{v_s}$ are not independent but intrinsically connected via the relation $A_{v_{\rm s}} = Mv_\mathrm{s}B_{v_{\rm s}}/k_\mathrm{B}T$. This guarantees that the collision integral vanishes when $f(v_{\rm s})$ is given by the classical Maxwell-Boltzmann distribution. For a positive mass object this corresponds to thermal equilibrium, however for negative mass particles the situation is more complicated.\\
\indent The crucial difference between equation~\eqref{eqn:collisionint} and the typical collision integral for a positive mass object is that the drift term $A_{v_{\rm s}} \propto v_{\rm s}$ is \emph{negative}, indicating that over time the distribution drifts from lower to higher velocities. Thus, equation~\eqref{eqn:kineticeqn} does not have a stationary solution. We show below that this leads to the absence of diffusion in free space, where diffusion is formally defined as variance in position that grows linearly in time, $\langle \bar{x}^2\rangle \propto t$. The apparent unbound runaway of the distribution function is a result of the expansion of the energy in equation~\eqref{eqn:energy} for $v_{\rm s} \ll c$, as discussed in further detail in Appendix C. If we consider the full energy spectrum then the system does reach equilibrium, where the dark soliton accelerates to the speed of sound and disappears. However, to describe the initial soliton trajectory we choose to work in the regime $v_{\rm s} \ll c$ where the collision integral takes the simple form given by equation~\eqref{eqn:collisionint}.\\
\indent We note that in the case of bosonic impurities, instead of Pauli blocking factor, there is a Bose enhancement factor, $\left[1+n_\mathrm{B}(\epsilon_{-k})\right]$ where $n_B$ is the Bose-Einstein distribution. This factor has been overlooked previously~\cite{Fedichev1999,Muryshev2002}, but it strongly influences the magnitude of the friction coefficient for a degenerate gas of impurities~\cite{Aycock2016}. Moreover, the factor is crucial for satisfying the fundamental relation $A_{v_{\rm s}} = Mv_\mathrm{s}B_{v_{\rm s}}/k_{\rm B}T$, which is dictated only by equilibrium properties and is not sensitive to the nature of the impurities~\cite{Lifschitz1983}.\\
\indent Finally, combining equations~\eqref{eqn:kineticeqn} and~\eqref{eqn:collisionint}, we find the following Kramer's type equation for the soliton distribution function 
\begin{equation}
\frac{\partial f}{\partial t}+v_{\rm s}\frac{\partial f}{\partial x_{\rm s}} = \frac{\partial}{\partial v_{\rm s}}\left(-\Gamma v_{\rm s}f - \frac{\partial_{x_{\rm s}}U}{M}f + \Gamma v_{\rm th}^2\frac{\partial f}{\partial v_{\rm s}}\right)
\label{eqn:Kramers}
\end{equation}
\begin{figure}[t!]
\centering
\includegraphics[width=\columnwidth]{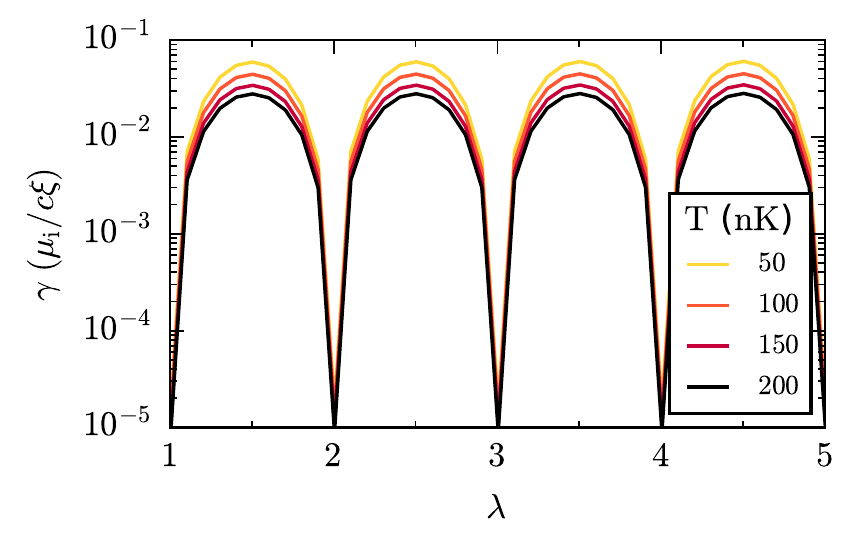}
\caption{(color online)~The soliton friction coefficient $\gamma$ is periodic as a function of $\lambda$ where $\lambda(\lambda-1) = 2m_\mathrm{i}g'/m_\mathrm{c}g$. Friction vanishes for integer $\lambda$, indicating that the soliton is reflectionless to scatterers. A system with tunable interactions enables tuning $\gamma$ without changing the number of scatterers. $\gamma$ is calculated in units of $\mu_{\rm i}/c\xi$ where $\mu_{\rm i}$ is the chemical potential of impurities, $c$ is the speed of sound in the condensate and $\xi$ is the condensate healing length. For impurities of $^{173}$Yb as we have calculated here, $\gamma$ decreases with increasing temperature. Increasingly dark lines indicate higher temperatures.
\label{Fig:gammavsL}}
\end{figure}
where $\Gamma = \gamma/M$ and $v^2_{\rm th} = k_\mathrm{B}T/M$. This equation is analytically solvable in the case of harmonic confinement, $U(x_{\rm s}) = -M_{\rm g}\omega_{\mathrm{c},x}^2x_{\rm s}^2/2$. We present the full solution for the distribution function $f(t,x_{\rm s},v_{\rm s})$ in Appendix C.\\
\indent The Langevin equation of motion for a single soliton can be inferred from equation~\eqref{eqn:Kramers}, and it is given by 
\begin{equation}
-M\ddot{x}_{\rm s} = -\gamma\dot{x}_{\rm s} + M\omega^2x_{\rm s} + f_{\rm s}(t)
\end{equation}
where $\omega = \omega_{\mathrm{c},x}/\sqrt{2}$. The stochastic Langevin force is characterized by white-noise correlations with $\langle f_{\rm s}(t) \rangle = 0$ and $\langle f_{\rm s}(t')f_{\rm s}(t)\rangle = 2\gamma k_{\rm B}T\delta(t-t')$. From this equation we see that $\gamma$ plays the role of the friction coefficient.\\
\indent At fixed impurity number, $\gamma$ depends on three parameters: the temperature $T$, the parameter $\lambda$, and the chemical potential $\mu$, which is itself a function of $\lambda$ and $T$. Figure~\ref{Fig:gammavsL} shows $\gamma$ as a function of $\lambda$ for four different temperatures; $\gamma$ grows over many orders of magnitude as $\lambda$ is tuned from integer to half integer values. The integral over $k$ in equation~\eqref{eqn:gamma} is strongly peaked around $k = 0$. Therefore, approximating $n_\mathrm{F}(\epsilon_{k= 0})\left[1-n_\mathrm{F}(\epsilon_{k = 0})\right] \approx e^{-\beta\mu_{\rm i}}$ and integrating over $k$, one finds the following expression for the friction coefficient, 
\begin{equation}
\gamma \approx \frac{3\hbar^3e^{-\mu_{\rm i}/k_{\rm B}T}}{2\pi^5 m_{\rm i}k_{\rm B}T\xi^4}\sin^2(\pi\lambda).
\label{eqn:gammaappx}
\end{equation}
Recalling that $\lambda(\lambda-1) = 2m_\mathrm{i}g'/m_\mathrm{c}g$, we see clearly that the system can be tuned to the frictionless limit where $\gamma = 0$ without changing the number of impurities. Furthermore, friction provides an anti-damping force to the soliton, while the background harmonic potential provides a confining force. The interplay of friction and confinement lead to the emergence of Brownian motion in the system.
\section{Dark Soliton Trajectory\label{Sec:Dark}} From the solution $f(t,x_{\rm s},v_{\rm s})$ of equation~\eqref{eqn:Kramers}, we can calculate exact expectation values for the soliton position and velocity. In experiments, it is generally easier to measure soliton position, which we focus on in the following. The average soliton trajectory is given by
\begin{equation}
\bar{x}_{\rm s}(t,\omega) = \frac{v_{\rm i}e^{\Gamma t/2}}{\bar{\omega}}\sin\left(\bar{\omega}t\right),
\label{eqn:pos}
\end{equation}
where $\bar{\omega} = \sqrt{\omega^2-\Gamma^2/4}$ and $v_{\rm i}$ is the initial velocity of the soliton. The variance in soliton position, $D_x$, is given by 
\begin{align}
D_x(t,\omega) &=\frac{v_{\rm th}^2(e^{\Gamma t}-1)}{\bar{\omega}^2} + \frac{v_{\rm i}^2e^{\Gamma t}}{\bar{\omega}^2}\sin^2\left(\bar{\omega}t\right)\label{eqn:posDiff}\\
&+\frac{v_{\rm th}^2\Gamma^2e^{\Gamma t}}{4\omega^2\bar{\omega}^2}\left[1-\left(\cos(2\bar{\omega}t)+\frac{2\bar{\omega}}{\Gamma}\sin(2\bar{\omega}t)\right)\right].\nonumber
\end{align}
The problem has an intrinsic timescale given by $\Gamma^{-1} = M/\gamma$. For $\Gamma t \gtrsim 1$, the soliton's position grows exponentially, indicative of the soliton rapidly reaching the speed of sound and disappearing. We examine equations~\eqref{eqn:pos} and~\eqref{eqn:posDiff} in the short-time limit $\Gamma t \ll 1$. The trap frequency $\omega$ also considerably affects the soliton dynamics. In the limit $\Gamma \ll \omega$, we find that diffusive behavior emerges, where $D_x(t,\omega) \propto D_0 + D(t)t$ with a time-dependent diffusion coefficient 
\begin{equation}
\langle D(t) \rangle \approx \frac{v_{\rm th}^2\Gamma}{\omega^2} + \frac{v_{\rm i}^2\Gamma}{\omega^2}\sin^2\left(\bar{\omega}t\right)-\frac{v_{\rm th}^2\Gamma^2}{2\omega^3}\sin(2\bar{\omega}t), 
\label{eqn:appxD}
\end{equation}
with offest $D_0 \approx v_{\rm i}^2/2\omega^2 + v_{\rm th}^2/4\omega^4$. However, in the opposite limit of $\omega \ll \Gamma$, the linear in $t$ term vanishes, giving 
\begin{equation}
D_x(t,\omega) \approx v_{\rm i}^2t^2 + v_i^2\Gamma t^3 + \frac{2}{3}v_{\rm th}^2\Gamma t^3
\end{equation}
\begin{figure}[th]
\centering
\includegraphics[width=\columnwidth]{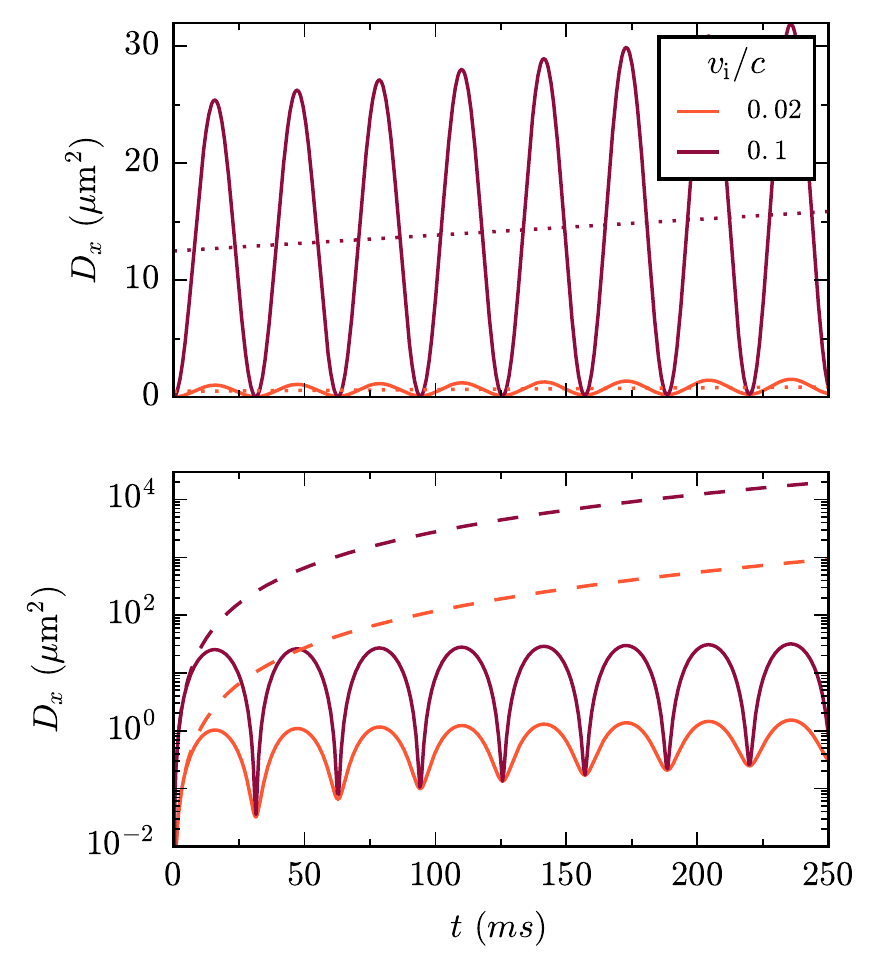}
\caption{(color online)~Variance in soliton position $D_x(t,\omega)$ as a function of time from the exact expression~\eqref{eqn:posDiff} for $v_{\rm i}/c = 0.02$ (orange/light gray line) and $v_{\rm i}/c = 0.1$ (black line) with $v_{\rm th} \approx 0.1~{\rm mm/s}$ and $\Gamma^{-1} \approx 1~{\rm s}$. Top: Results for a harmonic potential with $\omega = 100\Gamma$. $D_x$ grows linearly in time with additional oscillations due to confinement. The amplitude of oscillation increases with increasing $v_{\rm i}$. Dotted lines show the average value using the linear approximation in equation~\eqref{eqn:appxD}. Bottom: Comparison with $D_x(t,\omega)$ in the limit $\omega \ll \Gamma$ (dashed lines). In the absence of harmonic confinement, $D_x$ initially grows like $t^3$ for $\Gamma t \ll 1$, then grows exponentially. There is no diffusive regime.\label{Fig:Diffusion}}
\end{figure}
to lowest order in $\Gamma t$. In the absence of the restoring force provided by background potential, the soliton undergoes ballistic transport $\propto t^3$, followed by the exponential growth of $D_x$. The exact expression for $D_x(t,\omega)$ is shown in Figure~\ref{Fig:Diffusion}. The mechanism of diffusion for dark solitons is thus inherently different than Brownian motion for positive mass objects. Friction forces cause the soliton to speed up, therefore the only restoring force in the problem is due to the background confining potential, which leads to the emergence of diffusive behavior. Finally, we see that in the frictionless limit, $\Gamma \rightarrow 0$, we have $D \rightarrow 0$, and there is no diffusion. For quantitative agreement with experiment, the initial velocity of the soliton $v_{\rm i}$ also plays a crucial role~\cite{Aycock2016}.
\section{Soliton Lifetime\label{Sec:Soliton}} 
\begin{figure}[t!]
\centering
\includegraphics[width=\columnwidth]{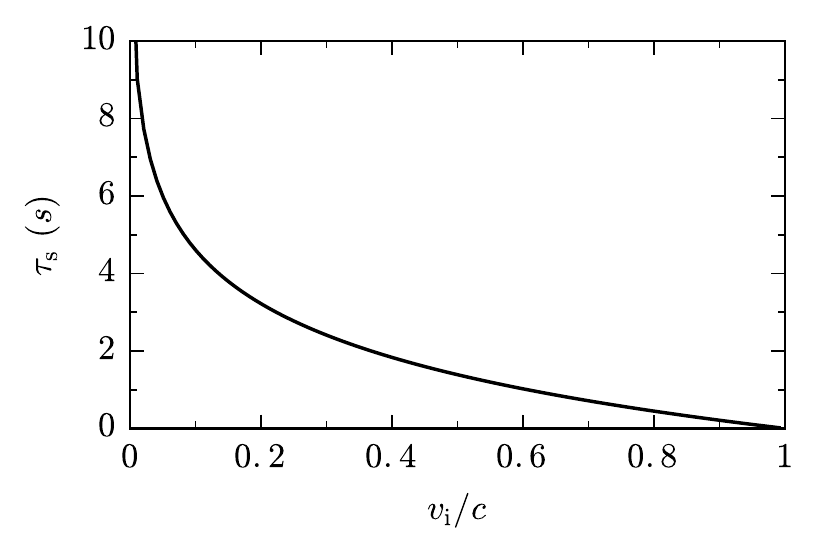}
\caption{Soliton lifetime as a function of initial velocity $v_{\rm i}$, with $\omega = 100\Gamma$ and where $c$ is the condensate speed of sound. Solitons that start at higher initial velocities have a shorter lifetime, which one would intuitively expect. Soliton lifetime is only weakly dependent on trapping frequency $\omega$.\label{Fig:Lifetime}}
\end{figure}

Integrating the distribution function $f(t,x_{\rm s},v_{\rm s})$ over the spatial coordinate $x_{\rm s}$, we find the distribution of soliton velocities 
\begin{equation}
f_v(t,v_{\rm s}) = \frac{1}{\sqrt{4\pi g_3(t,\omega)}}\exp\left(-\frac{(v_{\rm s}-\bar{v}_{\rm s}(t))^2}{4g_3(t,\omega)}\right),
\label{eqn:velocitypdf}
\end{equation}
parametrized by functions $g_3(t,\omega)$ and $\bar{v}_{\rm s}(t)$. The function $g_3(t,\omega)$ is given by 
\begin{equation}
g_3 = \frac{\left[4\bar{\omega}^2(e^{\Gamma t}-1)+e^{\Gamma t}\left(\Gamma^2+2\Gamma\bar{\omega}\sin(2\bar{\omega}t)-\Gamma^2\cos(2\bar{\omega}t)\right)\right]}{8\bar{\omega}^2}
\label{eqn:g31}
\end{equation}
The average velocity is given by $\bar{v}_{\rm s}(t)$, which captures oscillations in the trap as well as exponential growth of the soliton velocity,
\begin{equation}
\bar{v}_{\rm s}(t) = \frac{v_ie^{\Gamma t/2}}{\bar{\omega}}\left[\bar{\omega}\cos\left(\bar{\omega}t\right)+\frac{\Gamma}{2}\sin\left(\bar{\omega}t\right)\right].
\label{eqn:vbar}
\end{equation}
Starting from the velocity distribution function in equation~\eqref{eqn:velocitypdf}, we impose a perfectly absorbing boundary condition at $v_{\rm s} = \pm c$, reflecting that the soliton disappears once it reaches the speed of sound. This can be done using the method of images as discussed in detail in Appendix D. The total survival probability is defined by integrating over $f_v$ from $-c$ to $c$.  The final expression for survival probability is given by
\begin{equation}
\mathcal{P}(|v_{\rm s}|<c ; \tau) = \int_{-c}^{c}dv_{\rm s}f^{\rm Img}_v(t,v_{\rm s}),
\end{equation}
where $f^{\rm Img}_v(t,v_{\rm s})$ is the distribution function that obeys the boundary condition $f^{\rm Img}_v(t,\pm c) = 0$ for all $t$. The full expression for $\mathcal{P}(|v_{\rm s}|<c ; \tau)$ can be found in Appendix D. From the method of images construction, the soliton survival probability is exactly zero when $|\bar{v}_{\rm s}(t)| = c$. Using the maximum value of $\bar{v}_{\rm s}(t)$ over one period, given by $\bar{v}^*_{\rm s}(t) = v_{\rm i}\omega e^{\Gamma t/2}/\bar{\omega}$, we define the soliton lifetime as the time $\tau_{\rm s}$ where $\mathcal{P}(|v_{\rm s}|<c ; \tau_{\rm s}) = 0$ and $|\bar{v}^*_{\rm s}(\tau_{\rm s})| = c$. This gives a simple expression for the lifetime,
\begin{equation}
\tau_{\rm s} = \frac{2M}{\gamma}\ln\left(\frac{c~\bar{\omega}}{v_{\rm i}\omega}\right).
\label{eqn:lifetime}
\end{equation}
The soliton lifetime from equation~\eqref{eqn:lifetime} is shown in Figure~\ref{Fig:Lifetime}. The lifetime decreases as initial velocity increases, however it is only weakly dependent on the trapping frequency $\omega$. Furthermore, soliton lifetime is simply inversely proportional to the friction coefficient $\gamma$, and diverges as $\gamma \rightarrow 0$. Tuning the friction coefficient therefore should have a measurable effect in experiments, where soliton lifetime increases as $\gamma$ decreases.
\section{Discussion and Conclusion\label{Sec:Discussion}} We calculated the friction coefficient $\gamma$ and the diffusion coefficient $D(t)$ of a dark soliton in the presence of a non-interacting Fermi gas. We have shown that the soliton acts as a potential well for the fermionic impurities, and the scattering states and reflection coefficient of the impurities can be calculated exactly.\\
\indent In this section we estimate properties of a Bose-Fermi mixture of $^{174}\mathrm{Yb}-^{173}\mathrm{Yb}$, however the theory is general and applicable to other Bose-Fermi mixtures. We chose $^{174}\mathrm{Yb}-^{173}\mathrm{Yb}$ as a lab-realized example with scattering properties giving $m_{\rm i}g'/m_{\rm c}g\approx 1.3$~\cite{Kitagawa2008,Fukuhara2009}.\\
\indent We consider a quasi-1D BEC of $^{174}$Yb atoms with $n_{\rm c}\xi \approx 100$ and speed of sound $c\approx 1~\mathrm{mm}/\mathrm{s}$, corresponding to a soliton mass of $M \approx 600m_{\rm c}$ and chemical potential $\mu_{\rm c} \approx \hbar\times2~\rm{kHz}$. For $T = 150~{\rm nK}$, the thermal velocity of the soliton is $v_{\rm th} = \sqrt{k_{\rm B}T/M} \approx 0.1~{\rm mm/s}$. The chemical potential requires a radial trapping frequency of $\omega_{\mathrm{c}\perp} \gtrsim 2\pi\times 10~\mathrm{kHz}$ for the quasi-1D criterion to be satisfied, and the shallow trapping direction should have $\omega_{\mathrm{c},x} \lesssim 2\pi\times 100~{\rm Hz}$. We set the number of $^{173}$Yb impurity atoms to $N_\mathrm{i} = 1000$. We choose the system length $L = 250~\mu{\rm m}$, long enough that the continuum description of the impurity scattering states is appropriate. We find the chemical potential of fermions to be on the order of $k_\mathrm{B}T$, which requires a transverse trapping frequency $\omega_{\mathrm{i}\perp} \gtrsim 2\pi\times 10~\mathrm{kHz}$ for the fermions to be considered one-dimensional. For lower frequencies $\omega_{\mathrm{i}\perp}$ it is possible to obtain an accurate theory by summing over quantized transverse modes for the impurities~\cite{Aycock2016}.\\
\indent Changing the magnitude of friction for the soliton requires tunable interactions. The most straightforward way of tuning interactions in the Yb system is by changing the overlap of the transverse wavefunctions of the impurities and condensate atoms. This can be done by applying optical forces to either the bosonic or fermionic species which change the overlap of the atomic clouds. The narrow linewidths in the Yb spectra are ideal for this type of selective addressing, which can be done with high precision~\cite{Lemke2009}. Bose-Fermi mixtures with different atomic species allow for other ways of tuning interactions by using Feshbach resonances or magnetic field gradients~\cite{Kim2016}. For the Yb $s$-wave scattering lengths that have already been measured, we find $\lambda \approx 2.2$~\cite{Kitagawa2008}. Such a system would only need to be tuned to slightly weaker interactions such that $\lambda \approx 2$ to see a decrease in friction coefficient and corresponding measurable increase in soliton lifetime. For attractive interspecies interactions ($g'<0$) such as in $^{87}\mathrm{Rb}-^{40}\mathrm{K}$ the soliton appears as a potential barrier rather than a well. The theory is still applicable in this case; the form of the reflection coefficient $R(k)$ is slightly different but still periodic in $\lambda$ and the physics is nominally unchanged~\cite{Ccevik2016}.\\
\indent We developed a general theory for the stochastic dynamics of negative mass objects using a kinetic equation approach. We find that the dynamics are drastically different from their positive mass counterparts - they do not undergo Brownian motion in free space. The proposed dark soliton-Fermi gas system provides an ideal experimental testbed in which to further study how friction and dissipation affects and object with negative mass.\\
\indent We presented an analytical expression for the friction coefficient based on fermion scattering properties, including a term accounting for Pauli blocking, which is important to satisfy the equilibrium conditions on the transport coefficients. Using this result, we found exact expressions for the soliton position and position variance over time. We classified soliton trajectories at short times as diffusive and ballistic, and the diffusive regime can only be seen in the presence of a confining potential. The crossover timescale is given by $\Gamma^{-1} = M/\gamma$, which we find to be on the order of a second. The intrinsic frequency $\Gamma \sim 1\mathrm{Hz}$ is very low. Thus, the timescale over which diffusive behavior occurs is on the order of seconds and the diffusion coefficient can be directly measured~\cite{Aycock2016}. Furthermore, the limit $\Gamma \ll \omega$ is justified for a reasonably shallow trapping potential which still preserves the soliton shape. Experiments with tunable interspecies interaction strength present the ability to tune the amount of friction at fixed impurity number, providing a simple way to manipulate the lifetime and trajectory of dark solitons in a laboratory setting.
\section*{Acknowledgements}
This work was supported by US-ARO (contract No. W911NF1310172), NSF-DMR 1613029, and Simons Foundation (H.H. and V.G.). H.H. acknowledges additional fellowship support from the National Physical Science Consortium and NSA. Additional support was provided by the AROs atomtronics MURI, the AFOSRs Quantum Matter MURI, NIST, and the NSF through the PFC at the JQI (I.B.S.). Part of this work was completed at the Kavli Institute for Theoretical Physics (KITP) and the authors are grateful to KITP for hospitality and for the partial support of this research from the National Science Foundation under Grant No. NSF PHY-1125915 (H.H. and V.G.).
\section*{Appendix A: Chemical potential of fermions}
\indent In addition to the obvious dependence on $R(k,\lambda)$, the friction coefficient $\gamma$ is highly sensitive to the chemical potential of the fermionic atoms through the distribution function $n_\mathrm{F}(\epsilon_k)$ in equation~\eqref{eqn:gamma}. Although the number of fermions in the system is fixed, the chemical potential is sensitive to the bound states in the soliton well, the phase shift of the scattering states, and the density of states at the Fermi level. In this Appendix we present the full calculation of the chemical potential of 1D fermions in the presence of a dark soliton potential well.\\
\indent The total number of impurities is given by $N_\mathrm{i} = N_\mathrm{s}(\mu_\mathrm{i}) + N_\mathrm{b}(\mu_\mathrm{i}) + \delta N(\mu_\mathrm{i})$, where $N_\mathrm{s}$ indicates scattering (continuum) states, $N_\mathrm{b}$ indicates bound states, and $\delta N$ is a correction due to the phase shift of scattering states. All three quantities are a function of the chemical potential $\mu_\mathrm{i}$. We can define the following equation for the 1D impurity density,  
\begin{widetext}
\begin{equation}
   \frac{N_\mathrm{i}}{L} = -\sqrt{\frac{m_\mathrm{i}k_\mathrm{B}T}{2\pi\hbar^2}}\mathrm{Li}_{\frac{1}{2}}\left(e^{-\beta\mu_{\rm i}}\right)+\frac{2}{L}\sum_{j=0}^{\mathrm{floor}(\lambda-1)}\frac{1}{e^{\beta(\epsilon_j-\mu_{\rm i})}+1}
   +\frac{1}{L}\int\frac{dk}{2\pi}\frac{1}{e^{\beta(\epsilon_k -\mu_{\rm i})}+1}\frac{\partial\delta(k,\lambda)}{\partial k}\mbox{~~;~~}\beta = \frac{1}{k_\mathrm{B}T}
   \label{eqn:chempot}
\end{equation}
\end{widetext}
The first term in equation~\eqref{eqn:chempot} comes from integrating over $k$ for the continuum states, where $\mathrm{Li}_{1/2}(x)$ is the polylogarithm function. The continuum dispersion is $\epsilon_k \propto \hbar^2k^2/2m_\mathrm{i}$. The second term accounts for the bound states, which have quantized energies $\epsilon_j = -\hbar^2/2m_\mathrm{i}\xi^2(\lambda - 1-j)^2$ for integer $j <\lambda -1$. The factor of two accounts for Pauli degeneracy. Finally, the third term in equation~\eqref{eqn:chempot} is a correction due to the phase shift of the scattering states. The phase shift is given by $\delta(k,\lambda) = \mathrm{Arg}\left[t(k,\lambda)\right]$ with transmission amplitude
\begin{figure}[t!h]
\centering
\includegraphics[width=\columnwidth]{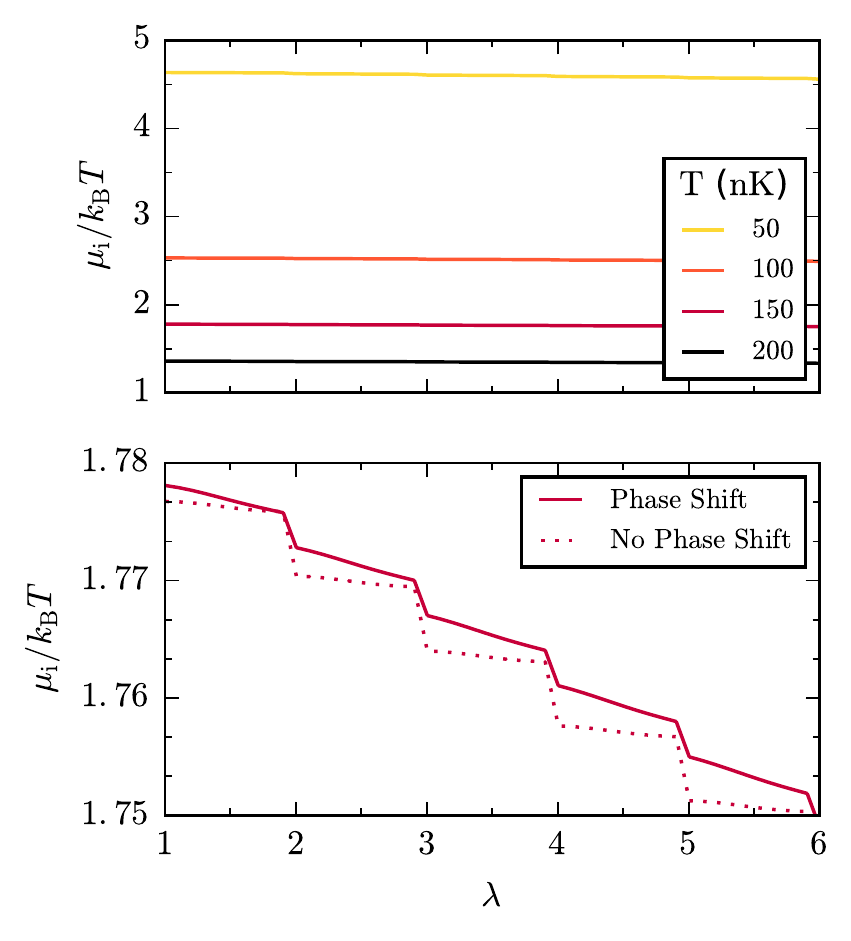}
\caption{(color online)~Top: The chemical potential $\mu_{\rm i}$ of fermionic impurities, from solving equation~\eqref{eqn:chempot} numerically for $N_{\rm i} = 1000$ $^{173}$Yb atoms with $L = 250~\mu{\rm m}$. $\mu_{\rm i}$ decreases slightly as $\lambda$ is increased and there are more bound states in the soliton well. Increasingly dark lines indicate higher temperatures. Bottom: Chemical potential for $T = 150~\mathrm{nK}$. The steps at each integer indicate an additional bound state in the soliton well. The chemical potential including the phase shift (solid line) is increased slightly from the result without it (dotted line).\label{Fig:allmu}}
\end{figure}
\begin{equation}
    t = \frac{\Gamma(\lambda-ik)\Gamma(1-\lambda-ik)}{\Gamma(1-ik)\Gamma(-ik)}, 
    \label{eqn:t}
\end{equation}
from the scattering matrix of equation~\eqref{eqn:PoschlTeller}. The correction $\delta N$ is proportional to $\partial_k\delta(k,\lambda)$, which takes the form
\begin{align}
    \frac{\partial\delta(k,\lambda)}{\partial k} &= \mathrm{Re}\left[\psi^0(-ik) + \psi^0(1-ik)\right.\nonumber\\
    &\left.- \psi^0(\lambda-ik) - \psi^0(1-ik-\lambda)\right],
    \label{eqn:Pshift}
\end{align}
where $\psi^0(z) = \Gamma'(z)/\Gamma(z)$ is the digamma function. We solve for $\mu_{\rm i}$ numerically for $^{173}$Yb atoms with $N_\mathrm{i} = 1000$ and $L = 250~\mu\mathrm{m}$ for temperatures $T =\left\{50,100,150,200\right\}~\mathrm{nK}$.\\
\indent The results of the calculation are shown in Figure~\ref{Fig:allmu}. The chemical potential of the fermion atoms decreases slightly as the interaction strengths are tuned and the soliton well becomes deeper, and we find $\mu_{\rm i} \approx k_\mathrm{B}T$ for all temperatures considered. In the bottom panel of Figure~\ref{Fig:allmu} we show the result with and without accounting for the phase shift term. The phase shift of the scattering states slightly increases the chemical potential. These results are used in calculating $\gamma$ in equation ~\eqref{eqn:gamma} and Fig.~\ref{Fig:gammavsL} of the main text.
\section*{Appendix B: Derivation of Collision Integral and Transport Coefficients}
The collision integral for a heavy object interacting with a gas of lighter objects can be derived in quite a general way, as has been done in many textbooks (e.g.\cite{Lifschitz1983,Risken1984}). The essential assumption is that the momentum transferred in each collision is small. Treating the soliton as a heavy classical object, we assume that it has some probability distribution $f(t,x_{\rm s},p)$ which depends on momentum $p$, time $t$, and position $x_{\rm s}$. Let $P(p,q)dq$ denote the probability per unit time of a change $p \rightarrow p-q$ in the momentum of the soliton in a collision with a fermionic impurity atom, where $q$ is the momentum transferred. The transport equation for $f$ is then given by
\begin{equation}
\frac{df}{dt} = \int dq ~ P(p+q,q)f(t,p+q) - P(p,q)f(t,p)
\end{equation}
which measures the difference between the soliton scattering into a state with momentum $p$ and out of a state with momentum $p$. We assume that the transferred momentum in each collision is small, i.e. $q \ll p$ and that $P(p,q)$ is a smooth function. We then make the following expansion, 
\begin{align}
P(p&+q,q)f(t,p+q)\approx P(p,q)f(t,p)\\ &+ q\frac{\partial}{\partial p}\left(P(p,q)f(t,p)\right) +\frac{1}{2}q^2\frac{\partial^2}{\partial p\partial p}\left(P(p,q)f(t,p)\right)\nonumber.
\end{align}
This gives a transport equation for $f$ with a collision integral in Fokker-Planck form 
\begin{equation}
\frac{df}{dt} = \mathcal{I}\left[f\right]\mbox{~~;~~} \mathcal{I}\left[f\right] = \frac{\partial}{\partial p}\left(A_pf + \frac{\partial}{\partial p}\left[B_pf\right]\right).
\label{eqn:Kinetic2}
\end{equation}
The transport coefficients are given by
\begin{equation}
A_p = \sum_q q P(p,q)\mbox{~~;~~}B_p = \frac{1}{2} \sum_q q^2 P(p,q),
\end{equation}
where $q$ is the momentum transferred between the heavy object and the light one in a single collision.\\
\indent For the dark soliton we write the momentum $p = -Mv_{\rm s}$, giving the collision integral as a function of velocity
\begin{align}
\mathcal{I}\left[f\right] &= \frac{\partial}{\partial v_{\rm s}}\left(-\frac{A_p}{M}f + \frac{\partial}{\partial v_{\rm s}}\left[\frac{B_p}{M^2}f\right]\right)\\
&=\frac{\partial}{\partial v_{\rm s}}\left(-A_{v_{\rm s}}f + \frac{\partial}{\partial v_{\rm s}}\left[B_{v_{\rm s}}f\right]\right).
\end{align}
The transport coefficients presented in equation~\eqref{eqn:A} and~\eqref{eqn:B} of the main text are related to $A_p$ and $B_p$ as $A_{v_{\rm s}} = A_p/M$ and $B_{v_{\rm s}} = B_p/M^2$; we present the rest of the calculation in terms of $v_{\rm s}$.\\
\indent Coefficients $A_{v_{\rm s}}$ and $B_{v_{\rm s}}$ are not independent. When $f = \exp(-E(v_{\rm s})/k_{\rm B}T)$ is the Maxwell-Boltzmann distribution the collision integral $\mathcal{I}\left[f\right]$ must vanish. For the soliton, we have $f(E(v_{\rm s})) \approx \exp(Mv_{\rm s}^2/2k_{\rm B}T)$. Plugging this into $\mathcal{I}\left[f\right]$, we find the relation 
\begin{equation}
\left(-A_{v_{\rm s}}+\frac{\partial B_{v_{\rm s}}}{\partial v_{\rm s}}\right)f +\frac{v_{\rm s}MB_{v_{\rm s}}}{k_{\rm B}T}f = 0.
\end{equation}
To first order, $\partial_{v_{\rm s}}B_{v_{\rm s}} = 0$, giving
\begin{equation}
A_{v_{\rm s}} = \frac{Mv_{\rm s}B_{v_{\rm s}}}{k{\rm_B}T}.
\label{eqn:AB}
\end{equation}
We emphasize that this relation relies only on the condition that $f(E(v_{\rm s}))$ is Maxwell-Boltzmann and not on the microscopic properties of the scatterers, and the microscopic expressions for $A_{v_{\rm s}}$ and $B_{v_{\rm s}}$ must satisfy this relation~\cite{Lifschitz1983}.\\
\indent The probability per unit time that the heavy object will undergo a scattering event is $P(p,q) = R_q|v_q| \mathbf{s}(k,k_{\mathrm{s}})$, where $R_q$ is the reflection coefficient, $|v_q|$ is the velocity and $\mathbf{s}(k,k_{\mathrm{s}})$ is a statistical factor which gives the occupation number of scatterers. In the case of bosonic impurities, $\mathbf{s}(k,k_{\mathrm{s}}) = n_{\rm B}(1+n_{\rm B})$, where $n_B$ is the Bose-Einstein distribution. This term accounts for bosonic enhancement.\\
\indent In the case of fermionic impurities, which we consider here, $\mathbf{s}(k,k_{\mathrm{s}}) = n_{\rm F}(1-n_{\rm F})$ where $n_F$ is the Fermi-Dirac distribution. This term accounts for Pauli blocking, which means that a fermion with momentum $k$ is unable to scatter into a state with momentum $-k$ if that state is already filled. An incoming particle with momentum $p_{\rm i} = \hbar k$ is reflected with momentum $p_{\rm f} = -\hbar k$, giving $q = p_{\rm f} - p_{\rm i} = -2\hbar k$. Now, we have the following expressions for $A_{v_{\rm s}}$ and $B_{v_{\rm s}}$ 
\begin{align}
A_{v_{\rm s}} &= -\frac{2\hbar}{M}\sum_k k R_{k,\lambda}\left|\frac{\partial \epsilon_k}{\hbar \partial k}\right| n_\mathrm{F}(\epsilon_{k+k_\mathrm{s}})\left[1-n_\mathrm{F}(\epsilon_{-k+k_\mathrm{s}})\right].
\label{eqn:A2} \\
B_{v_{\rm s}} &= \frac{2\hbar^2}{M^2}\sum_k k^2R_{k,\lambda}\left|\frac{\partial \epsilon_k}{\hbar \partial k}\right| n_\mathrm{F}(\epsilon_{k+k_\mathrm{s}})\left[1-n_\mathrm{F}(\epsilon_{-k+k_\mathrm{s}})\right].\label{eqn:B2}
\end{align}
Given these relations, we can check that equation~\eqref{eqn:AB} is satisfied to first order in $k_s = m_{\rm i}v_s/\hbar$. The Fermi-Dirac distribution can be expanded as 
\begin{equation}
n_\mathrm{F}(\epsilon_{\pm k + k_{\rm s}}) \approx n_\mathrm{F}(\epsilon_{\pm k}) \pm \hbar k v_{\rm s}\frac{\partial n_{\rm F}}{\partial \epsilon_{\pm k}},  
\end{equation}
where we have used the relation $\epsilon_{\pm k+k_{\rm s}}\approx \epsilon_{\pm k}\pm \hbar k v_{\rm s}$. Plugging into equation~\eqref{eqn:A2}, we find 
\begin{align}
A_{v_{\rm s}} &\approx -\frac{2\hbar^2v_{\rm s}}{M}\sum_k k^2 R_{k,\lambda}\left|\frac{\partial \epsilon_k}{\hbar \partial k}\right|\frac{\partial n_{F}}{\partial\epsilon_k}\nonumber\\
%&\approx \frac{2\hbar^2v_{\rm s}}{Mk_{\rm B}T}\sum_k k^2 R_{k,\lambda}\left|\frac{\partial \epsilon_k}{\hbar \partial k}\right|n_{\rm F}(\epsilon_k)\left[1-n_{\rm F}(\epsilon_{-k})\right]\nonumber\\ 
&=\frac{Mv_{\rm s}B_{v_{\rm s}}}{k_{\rm B}T}
\end{align}
We note that the zeroth-order term of $A_{v_{\rm s}}$ vanishes because $\sum_k k$ is an odd function of $k$. Similarly, for $B_{v_{\rm s}}$ the first order in $k_{\rm s}$ is $B \propto \sum_k k^3 = 0$. Equation~\eqref{eqn:AB} is satisfied \emph{only} if the Pauli-blocking term is included in the microscopic expressions for $A_{v_{\rm s}}$ and $B_{v_{\rm s}}$.  
\section*{Appendix C: Solution of the Kinetic Equation for a soliton in Harmonic Trap}
Here we present an analytical solution of equation~\eqref{eqn:Kramers} for a soliton in harmonic trap with potential $U(x)=- M\omega^2 x^2/2$. In the following it is instructive to introduce dimensionless units as follows $t \rightarrow t/\Gamma$, $\omega \rightarrow \omega\Gamma$, $v_s\rightarrow v_{\rm th} v_s$, $x\rightarrow v_{\rm th}x/\Gamma$. The kinetic equation is given by
\begin{equation}
\frac{\partial f}{\partial t} + v_{\rm s}\frac{\partial f}{\partial x_{\rm s}} - \omega^2 x_\mathrm{s} \frac{\partial f}{\partial v_\mathrm{s}}= \frac{\partial }{\partial v_{\rm s}}\left(-v_{\rm s} f + \frac{\partial f}{\partial v_{\rm s}}\right), 
\label{eqn:KineticRealUnits}
\end{equation}
where $\Gamma = \gamma/M$ and $v^2_{\rm th} = k_\mathrm{B}T/M$ can be interpreted as the thermal velocity. Equation~\eqref{eqn:KineticRealUnits} needs to be supplemented by the initial conditions. We assume that the soliton is created in the trap center with initial velocity $v_i$, resulting in $f(0,x_{\rm s},v_{\rm s}) = \delta(x_{\rm s})\delta(v_{\rm s} - v_{\rm i})$. This second-order partial differential equation (PDE) can be reduced to first-order by Fourier transform. Setting 
\begin{equation}
f(t,x_{\rm s},v_{\rm s}) = \sum_{p,q}\bar{f}(t,p,q)e^{ipx_{\rm s}+iqv_{\rm s}},
\label{eqn:FourierK}
\end{equation}
we find the following first-order PDE 
\begin{equation}
\frac{\partial \bar{f}}{\partial t} -(p+q)\frac{\partial \bar{f}}{\partial q} + \omega^2q\frac{\partial \bar{f}}{\partial p} = -q^2\bar{f}
\label{eqn:qDiff}
\end{equation}
with the transformed initial condition $\tilde{f}(0,p,q) = e^{-iqv_i}$. Equation~\eqref{eqn:qDiff} can be solved analytically using the method of characteristics. According to the method, the PDE can be transformed to a system of ordinary differential equations (ODE) along characteristic lines, parametrized by $s$ and defined as follows
\begin{widetext}
\begin{align}
\label{eqn:Characteristic1}
\frac{d\bar{f}'}{ds} &= -q^2(s)\bar{f}, \quad &\frac{dt}{ds}= 1 \quad 
&\frac{dq}{ds} = -\left[p(s) + q(s)\right] \quad  &\frac{dp}{ds} = \omega^2q(s) \\
\bar{f}'(0) &=\bar{f}(0,p_0,q_0), \quad &t(0) = 0, \quad &q(0) = q_0, \quad &p(0) = p_0
\label{eqn:Characteristic1b},
\end{align}
\end{widetext}
where we introduced $\bar{f}'(s)=f(t(s),p(s),q(s))$ and equations~\eqref{eqn:Characteristic1b} are the initial conditions written in a general form. Integration of~\eqref{eqn:Characteristic1} results in 
\begin{widetext}
\begin{align}
&\bar{f}(s)= \exp\left(Z\right)\\
&Z= \frac{e^{-s}}{8\omega^2\bar{\omega}^2}\left[4\omega^2p_0(p_0+q_0) + 4q_0^2\omega^4 - (p_0^2 + 4p_0q_0\omega^2 + q_0^2\omega^2)\cos(2\bar{\omega}s) + 2\bar{\omega}(p_0^2-q_0^2\omega^2)\sin(2\bar{\omega}s)\right] -\frac{q_0\omega^2+p_0^2}{2\omega^2} - iq_0v_i\nonumber\\
&t(s) = s \quad p(s) =\frac{e^{-s/2}}{\bar{\omega}}\left[p_0\bar{\omega}\cos\left(\bar{\omega}s\right) + \frac{p_0 + 2q_0\omega^2}{2}\sin\left(\bar{\omega}s\right)\right]\quad
q(s) =\frac{e^{-s/2}}{\bar{\omega}}\left[q_0\bar{\omega}\cos\left(\bar{\omega}s\right) - \frac{2p_0 + q_0}{2}\sin\left(\bar{\omega}s\right)\right]
\label{eqn:Characteristic2}
\end{align}
\end{widetext}
Where $\bar{\omega} = \sqrt{\omega^2-1/4}$. For each initial point $q_0$, $p_0$ and the parameter $s$ we have the corresponding point $t(s,q_0,p_0)$, $q(s, q_0, p_0)$, $p(s, q_0, p_0)$ on the characteristic line along with $f'(s)$. After inversion of these relations $s(t,q,p)$, $q_0(t,q,p)$, $p_0(t,q,p)$ the general form of solution of the kinetic equation in the Fourier space is $\bar{f}(t,q,p)=\bar{f}'(s(t,q,p))$. The inversion of equations~\eqref{eqn:Characteristic2} leads to 
\begin{widetext}
\begin{align}
s = t, \quad \quad
p_0 = \frac{e^{t/2}}{\bar{\omega}}\left[p\bar{\omega}\cos\left(\bar{\omega}t\right)-\frac{p+2q\omega^2}{2}\sin\left(\bar{\omega}t\right)\right], \quad 
q_0 = \frac{e^{t/2}}{\bar{\omega}}\left[q\bar{\omega}\cos\left(\bar{\omega}t\right)+\frac{2p+q}{2}\sin\left(\bar{\omega}t\right)\right].
\label{eqn:char3}
\end{align}
\end{widetext}
Where equations~\eqref{eqn:char3} satisfy initial conditions~\eqref{eqn:Characteristic1b}. Finally, the general solution in Fourier space is given by 
\begin{widetext}
\begin{equation}
\tilde{f}(t,p,q) = \exp\left\lbrace-g_1(t,\omega)p^2 + g_2(t,\omega)pq - g_3(t,\omega)q^2 - iv_{\rm i}\left(g_4(t,\omega)p+g_5(t,\omega)q\right)\right\rbrace
\label{eqn:qSpacef}
\end{equation}
\end{widetext}
parametrized by the functions $g_i(t,\omega)$, where 
\begin{align}
g_1(t,\omega) &= \frac{4\bar{\omega}^2(e^t-1)-e^t\left[\cos(2\bar{\omega}t)+2\bar{\omega}\sin(2\bar{\omega}t)-1\right]}{8\omega^2\bar{\omega}^2}\label{eqn:g1}\\
g_2(t,\omega) &= \frac{e^t}{2\bar{\omega}^2}\left[\cos(2\bar{\omega}t)-1\right]
\end{align}
\begin{align}
g_3(t,\omega) &= \frac{4\bar{\omega}^2(e^t-1)+e^t\left[2\bar{\omega}\sin(2\bar{\omega}t)-\cos(\bar{\omega}t)+1\right]}{8\bar{\omega}^2}\label{eqn:g3}\\
g_4(t,\omega) &= \frac{e^{t/2}}{\bar{\omega}}\sin\left(\bar{\omega}t\right)\\
g_5(t,\omega) &= \frac{e^{t/2}}{\bar{\omega}}\left[\bar{\omega}\cos\left(\bar{\omega}t\right)+\frac{1}{2}\sin\left(\bar{\omega}t\right)\right]\label{eqn:g5}.
\end{align}
Equation~\eqref{eqn:qDiff} does not have a stationary solution where $\partial \tilde{f}/\partial t \rightarrow 0$, due to negative drift term, which causes the distribution to drift to higher velocities. Equation~\eqref{eqn:qSpacef} grows exponentially at long times; this is an artifact of the linear approximation for soliton momentum, $p \approx -Mv$. The approach is equally valid for the full soliton spectrum in equation~\eqref{eqn:energy}, which is bounded, but does not admit an exact analytical solution. Finally, transforming equation~\eqref{eqn:qSpacef} back to real space we find the full distribution function $f(t,x_{\rm s},v_{\rm s})$ with Gaussian form
\begin{widetext}
\begin{equation}
\begin{split}
f(t,x_{\rm s},v_{\rm s}) &= \frac{1}{2\pi\sqrt{4g_1g_3-g_2^2}}\exp\left\lbrace-\frac{1}{4g_1g_3-g_2^2}\left[g_1v_{\rm s}^2+g_3x_{\rm s}^2 +g_2v_{\rm s}x_{\rm s}\right.\right. \\ 
&\left. \left. -v_{\rm i}v_{\rm s}(g_2g_4+2g_1g_5) - v_{\rm i}x_{\rm s}(g_2g_5+2g_3g_4)+v_{\rm i}^2(g_3g_4^2+g_1g_5^2+g_2g_4g_5)\right]\right\rbrace.
\label{eqn:RealSpacef}
\end{split}
\end{equation}
\end{widetext}
The distribution contains all information about the stochastic dynamics of a  dark soliton. The time dependence of the average soliton position and variance in position are given by
\begin{widetext}
\begin{align}
    \bar{x}_{\rm s}(t) &= \int_{-\infty}^{\infty} dv_{\rm s} \int_{-\infty}^{\infty} dx_{\rm s}~x_{\rm s}~f(t,x_{\rm s},v_{\rm s})  = v_{\rm i}g_4(t,\omega)\rightarrow\frac{v_{\rm i}e^{\Gamma t/2}}{\bar{\omega}}\sin\left(\bar{\omega}t\right)\\
    D_x(t) &=   \int_{-\infty}^{\infty} dv_{\rm s} \int_{-\infty}^{\infty} dx_{\rm s}~x^2_{\rm s}~f(t,x_{\rm s},v_{\rm s}) = 2g_1(t,\omega) + v_{\rm i}^2g^2_4(t,\omega) \rightarrow \nonumber\\
    &\rightarrow \frac{v_{\rm i}^2e^{\Gamma t}}{\bar{\omega}^2}\sin^2\left(\bar{\omega}t\right)+\frac{v_{\rm th}^2(e^{\Gamma t}-1)}{\bar{\omega}^2}-\frac{v_{\rm th}^2\Gamma^2e^{\Gamma t}}{4\omega^2\bar{\omega}^2}\left(\cos(2\bar{\omega}t)+\frac{2\bar{\omega}}{\Gamma}\sin(2\bar{\omega}t)-1\right).
\end{align}
\end{widetext}
\begin{figure*}[htbp]
\centering
\includegraphics[width=\textwidth]{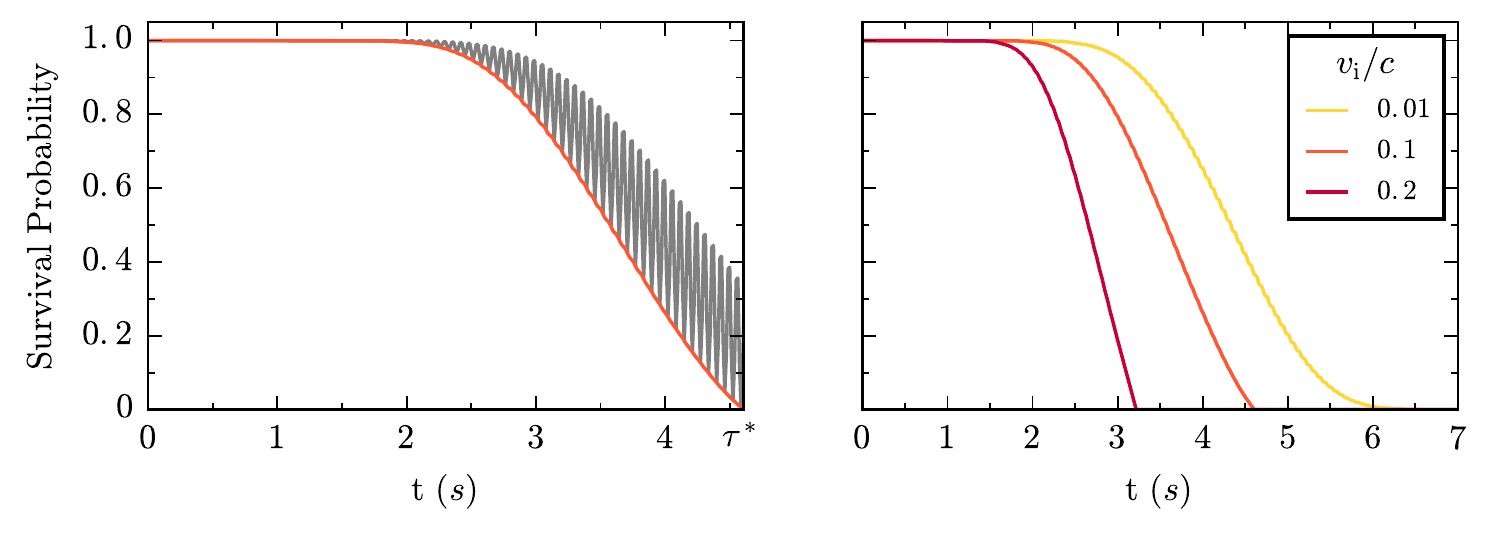}
\caption{(color online)~Left: Survival probability as defined by the exact expression in equation~\eqref{eqn:survivalProb} (gray line) is highly oscillatory due to the dynamics in the trap. The lower bound is given by replacing $\bar{v}_{\rm s}(t)$ with equation~\eqref{eqn:vbarmax} (black line). Soliton lifetime is marked by $\tau_{\rm s}$ on the horizontal axis. Calculated for $v_{\rm i} = 0.1c$ and $\omega= 50\Gamma$. Right: Survival probability for different soliton initial velocities with $\omega = 50\Gamma$. Survival probability falls off more quickly for faster initial velocities, with the fastest initial velocity indicated by the red (dark gray) line. \label{Fig:Survival}}
\end{figure*}
Where $\bar{\omega} = \sqrt{\omega^2-\Gamma^2/4}$ in real units. Similarly, the average velocity and variance in velocity are given by
\begin{widetext}
\begin{align}
    \bar{v}_{\rm s}(t) &= \int_{-\infty}^{\infty} dv_{\rm s} \int_{-\infty}^{\infty} dx_{\rm s}~v_{\rm s}~f(t,x_{\rm s},v_{\rm s})  = v_{\rm i}g_5(t,\omega) \rightarrow \frac{v_{\rm i}e^{\Gamma t/2}}{\bar{\omega}}\left[\bar{\omega}\cos\left(\bar{\omega}t\right)+\frac{\Gamma}{2}\sin\left(\bar{\omega}t\right)\right],\label{eqn:barv}\\
    D_v(t) &=   \int_{-\infty}^{\infty} dv_{\rm s} \int_{-\infty}^{\infty} dx_{\rm s}~v^2_{\rm s}~f(t,x_{\rm s},v_{\rm s}) \nonumber = 2g_3(t,\omega) + v_{\rm i}^2g_5(t,\omega)^2\rightarrow \nonumber \\
    &\rightarrow \frac{v_{\rm i}^2e^{\Gamma t}}{\bar{\omega}^2}\left[\bar{\omega}\cos\left(\bar{\omega}t\right)+\frac{\Gamma}{2}\sin\left(\bar{\omega}t\right)\right]^2+v_{\rm th}^2(e^{\Gamma t}-1)+\frac{\Gamma^2v_{\rm th}^2e^{\Gamma t}}{4\bar{\omega}^2}\left(\frac{2\bar{\omega}}{\Gamma}\sin(2\bar{\omega}t)-\cos(2\bar{\omega}t)+1\right).
\end{align}
\end{widetext}
\section*{Appendix D: Calculation of Soliton Lifetime}
The lifetime of a dark soliton depends on it's velocity; when the soliton reaches the condensate speed of sound $c$, its depth is zero and it disappears. Integrating equation~\eqref{eqn:RealSpacef} over spatial coordinate $x_{\rm s}$, we find the distribution function only in terms of velocity.
\begin{equation}
f_v(t,v_{\rm s}) = \frac{1}{\sqrt{4\pi g_3(t,\omega)}}\exp\left(-\frac{(v_{\rm s}-\bar{v}_{\rm s}(t))^2}{4g_3(t,\omega)}\right)
\label{eqn:velocitydistf}
\end{equation}
Where $g_3(t,\omega)$ is given by equation~\eqref{eqn:g3} and $\bar{v}_{\rm s}(t)$ is equation~\eqref{eqn:barv} in Appendix C.\\
\indent We consider a perfectly absorbing boundary condition at $v_{\rm s} = \pm c$ which accounts for the soliton disappearance. The boundary condition imposed is $f_v(\pm c,t) = 0$ for all times $t$. We construct a distribution that obeys this boundary condition using the method of images. The distribution function is reflected about the boundaries, $v_{\rm s} = \pm c$ in our case, with ``image distributions" placed at $v_n = 2cn$ for $n=\pm 1, \pm 2, ...$. The distribution which obeys the boundary condition is then described by the general formula
\begin{align}
f^{\rm Img}_v(t,v_{\rm s}) &= f_v(t,v_{\rm s}) + \sum_{n = 1}^{\infty}(-1)^n\left[f_v(t,v_n- v)\right.\nonumber\\
&\left.+f_v(t,-v_n - v)\right]\mbox{~~;~~}v_n = 2cn.
\end{align}
We find the total survival probability of the soliton by integrating over $v_{\rm s}$ from $-c$ to $c$,
\begin{equation}
\mathcal{P}(|v_{\rm s}| < c ; t) = \int_{-c}^{c} dv_{\rm s}~f^{\rm Img}_{v_{\rm s}}(v_{\rm s},t). 
\end{equation}
Since each term in $f^{\rm Img}_{v_{\rm s}}(v_{\rm s},t)$  is a Gaussian, we integrate and find the following expression for the survival probability:
\begin{widetext}
\begin{align}
\mathcal{P}(|v_{\rm s}| < c ; t) &= \frac{1}{2}\left(\mathrm{Erf}\left[\frac{c-\bar{v}_{\rm s}(t)}{2\sqrt{g_3(t,\omega)}}\right] + \mathrm{Erf}\left[\frac{c+\bar{v}_{\rm s}(t)}{2\sqrt{g_3(t,\omega)}}\right]\right) +\frac{1}{2}\sum_{n=1}^{\infty}(-1)^n\left(\mathrm{Erf}\left[\frac{c+v_n-\bar{v}_{\rm s}(t)}{2\sqrt{g_3(t,\omega)}}\right]\right.\nonumber\\
&\left.+ \mathrm{Erf}\left[\frac{c-v_n+\bar{v}_{\rm s}(t)}{2\sqrt{g_3(t,\omega)}}\right] +\mathrm{Erf}\left[\frac{c-v_n-\bar{v}_{\rm s}(t)}{2\sqrt{g_3(t,\omega)}}\right] + \mathrm{Erf}\left[\frac{c+v_n+\bar{v}_{\rm s}(t)}{2\sqrt{g_3(t,\omega)}}\right]\right)
\label{eqn:survivalProb}
\end{align}
\end{widetext}
The exact expression in equation~\eqref{eqn:survivalProb} is oscillatory, because $\bar{v}_{\rm s}(t)$ and $g_3(t,\omega)$ capture oscillations in the trap as well as the long-time acceleration of the soliton. However, if the soliton velocity reaches $c$ at any point in it's oscillation it will not survive, so we focus on the lower bound of $\mathcal{P}(|v_{\rm s}| < c ; t)$, which can be found by replacing $\bar{v}_{\rm s}(t) \rightarrow \bar{v}^*_{\rm s}(t)$ where $\bar{v}^*_{\rm s}(t)$ is the maximum value over one oscillation period,
\begin{equation}
\bar{v}^*_{\rm s}(t) = \frac{v_{\rm i}e^{\Gamma t/2}}{\bar{\omega}}\sqrt{\bar{\omega}^2 + \Gamma^2/4} = \frac{v_{\rm i}\omega e^{\Gamma t/2}}{\bar{\omega}}.
\label{eqn:vbarmax}
\end{equation}
Making this substitution, we can plot a smooth survival probability curve. When $|\bar{v}^*_{\rm s}(\tau_{\rm s})| = c$, $\mathcal{P}(|v_{\rm s}| < c ; \tau_{\rm s}) = 0$, and we define $\tau_{\rm s}$ as the soliton lifetime. The exact oscillatory expression and the lower envelope of the survival probability are shown in Figure~\ref{Fig:Survival}.
\bibliography{main}
\end{document}